\documentclass{iau}
\usepackage{graphicx}

\def\be{\begin{equation}}
\def\ee{\end{equation}}

\title[Outflow from hot accretion flows] 
{Outflow from Hot Accretion Flows}

\author[Feng Yuan, Defu Bu \& Maochun Wu]   
{Feng Yuan$^1$, Defu Bu$^1$, \and  Maochun Wu$^1$}

\affiliation{$^1$ Shanghai Astronomical Observatory, CAS; Shanghai 200030, China \\ email: {\tt fyuan@shao.ac.cn}}

\pubyear{2012}
\volume{290}  
\jname{Feeding compact objects: Accretion on all scales}
\editors{C.M. Zhang, T. Belloni, M. M\'endez \& S.N. Zhang, eds.}

\begin{document}

\maketitle

\begin{abstract}
Numerical simulations of hot accretion flows have shown that the mass accretion rate decreases with decreasing radius. Two models have been proposed to explain this result. In the adiabatic inflow-outflow solution (ADIOS), it is thought to be due to the loss of gas in outflows. In the convection-dominated accretion flow (CDAF) model, it is explained as because that the gas is locked in convective eddies. In this paper we use hydrodynamical (HD) and magnetohydrodynamical (MHD) simulations to investigate which one is physical. We calculate and compare various properties of inflow (gas with an inward  velocity) and outflow (gas with an outward velocity). Systematic and significant differences are found. For example, for HD flows, the temperature of outflow is  higher than inflow; while for MHD flows, the specific angular momentum of outflow is much higher than inflow. We have also analyzed the convective stability of MHD accretion flow and found that they are stable. These results suggest that systematic inward and outward motion must exist, i.e., the ADIOS model is favored. The different properties of inflow and outflow also suggest that the mechanisms of producing outflow in HD and MHD flows are buoyancy associated with the convection and the centrifugal force associated with the angular momentum transport mediated by the magnetic field, respectively. The latter mechanism is similar to the Blandford \& Payne mechanism but no large-scale open magnetic field is required here. Possible observational applications are briefly discussed.

\end{abstract}

\firstsection 
\section{Introduction}

In the early analytical studies of hot accretion flow such as ADAFs, it was  assumed that the mass accretion rate is a constant of radius. Correspondingly, the density follows $\rho (r) \propto r^{-3/2}$. Later many numerical simulations have been performed, including both hydrodynamical (HD) and magneto-hydrodynamical (MHD) ones (e.g., Igumenshchev \& Abramowicz 1999; Stone, Pringle \& Begelman 1999; Stone \& Pringle 2001). One of the most important findings is that the inflow and outflow rates (defined as the mass flux of flow with an inward and outward radial velocity) decrease with decreasing radius. These simulations, no matter they are HD and MHD ones, indicate that the profiles of inflow rate are quite similar, which can be described by, \be \dot{M}_{\rm in}(r) = 2\pi r^{2} \int_{0}^{\pi} \rho \min(v_{r},0)
   \sin \theta d\theta = \dot{M}_{\rm in}(r_{\rm out})\left(\frac{r}{r_{\rm out}}\right)^{s},\ee with $s\sim 0.5-1$ (see a review in Yuan, Wu \& Bu 2012). Correspondingly, the radial profile of density becomes flatter, $\rho(r)\propto r^{-p}$ with $p\sim 1-0.5$. The consistency between HD and MHD results is surprising given that the mechanisms of producing the inflow rate profile in  HD and MHD cases are different, as we will illustrate in \S2.3. But such a result is what predicted in the recent work of Begelman (2012). We want to emphasize that these theoretical results have been confirmed by observations to, e.g., Sgr A* and NGC~3115 (see Yuan, Wu \& Bu 2012 for details).

A natural question is then: what is the nature of the inward decrease of the inflow rate? Two models have been proposed. One is the adiabatic inflow-outflow solution (ADIOS; Blandford \& Begelman 1999; 2004; Begelman 2012). In this model the inward decrease of inflow rate is because of the mass loss in the outflow launched at every radius. The second model is the convection-dominated accretion flows (CDAFs; Narayan, Igumenshchev, \& Abramowicz 2000; Quataert \& Gruzinov 2000). This model is based on the assumption that the hot accretion flows, both HD and MHD, are convectively unstable. In this scenario, the inward decrease of mass accretion rate is because that with accretion, more and more fluid is locked in convective eddies operating circular motion.

Answering this question is the first main of the work by Yuan, Bu \& Wu (2012). In that work, we have investigated this problem from the following two aspects using HD and MHD numerical simulations. We first compare the various properties of both inflow and outflow, including the radial and rotational velocities, temperature, and Bernoulli parameter. If the CDAF scenario is correct, i.e, the motion of the flow is dominated by convective turbulence, and correspondingly inflow and outflow rates are simply due to turbulent fluctuation, we should expect that the properties of inflow and outflow are almost the same. As we will see, however, we find that the properties of inflow and outflow are systematically and significantly different (\S2.1). The second aspect is to directly analyze the convective stability of an MHD accretion flow. The HD hot accretion flow is convectively unstable. However, there has been a debate on the convective stability of an MHD flow. Yuan, Bu \& Wu (2012) used the simulation data to analyze the convective stability of MHD accretion flows and finds that the flow is stable (\S2.2). Based on the above results, they have concluded that the decrease of accretion rate for both HD and MHD flows is not because of convection, but systematic outflow. The origin of outflow and the observational applications are also discussed in \S2.3 and \S3, respectively.

\section{HD and MHD Numerical Simulations: ADIOS or CDAF?}
We have performed both HD (Model A) and MHD (Model B) simulations of two-dimensional axisymmetric accretion flows around black holes using the ZEUS code (Yuan, Bu \& Wu 2012).  The initial condition of the two models is identical, which is a rotating torus with constant angular momentum. The difference is that magnetic field is included in Model B, which is confined to the interior of the torus. 

\subsection{Different properties of inflow and outflow}

We have calculated the angle-integrated and mass flux-weighted value of some quantities. The results are shown in Figs. 1-3. From the figures, we can clearly see that the properties of inflow and outflow are significantly different. 
For example, we see from Fig. 1 that for Model A, the temperature of outflow is significantly higher than that of inflow. From Fig. 2 we see that for both Model A and B, the radial profile of the radial velocity (in unit of Keplerian velocity) is quite different for inflow and outflow, with the former increasing rapidly inward while the latter keeping roughly constant. From Fig. 3 we see that for Model B, the outflow has much larger angular momentum than the inflow. The difference between Model A and B also suggests that the mechanisms of producing outflow in the HD and MHD cases are different, as we will discuss in \S2.3. This indicates that the inflow and outflow are not simply due to convective (for Model A) and MHD (for Model B) turbulence. Rather, they are systematic inflowing and outflowing motion.

\begin{figure}
\begin{center}
 \includegraphics[width=2.4in]{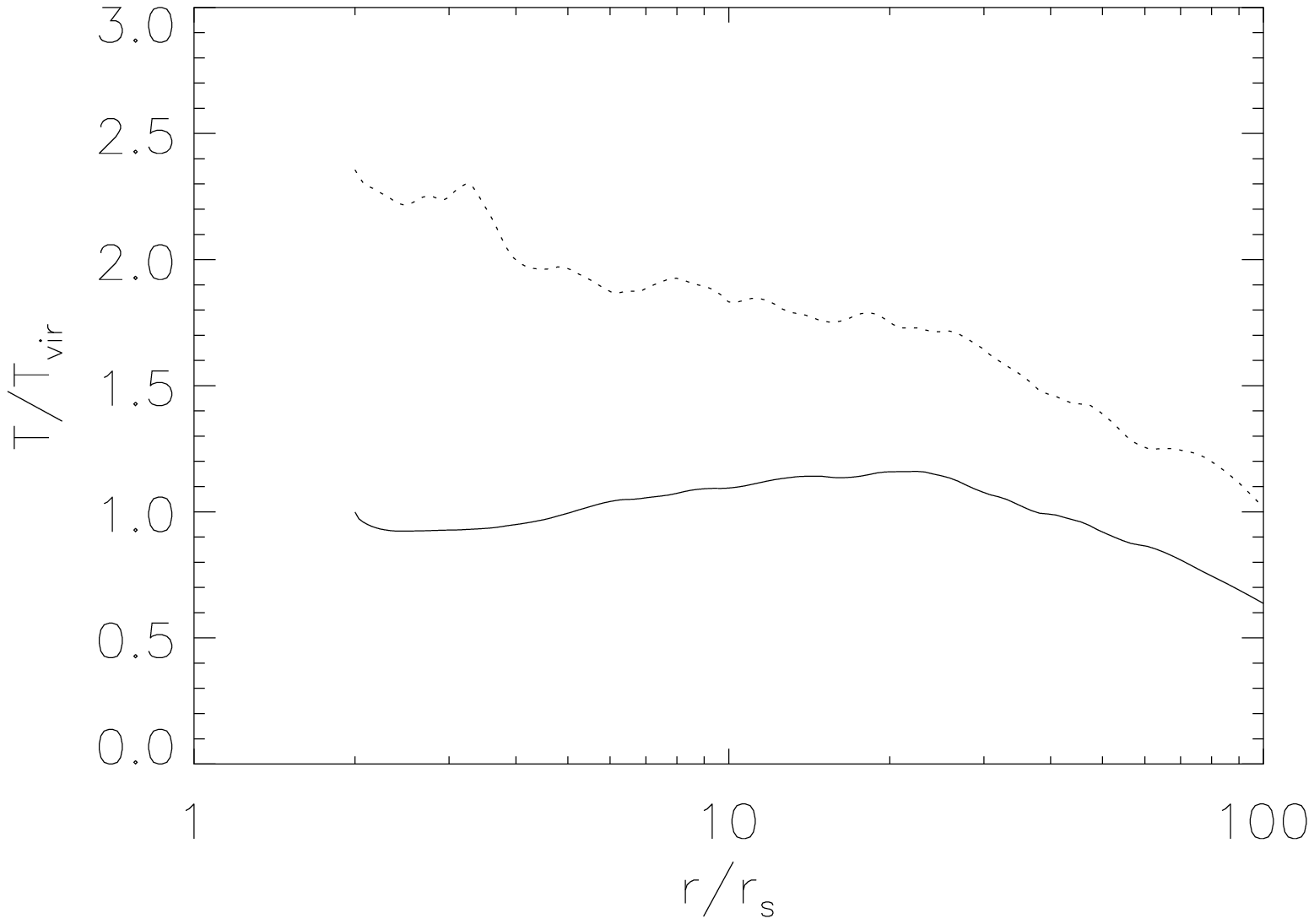}
 \includegraphics[width=2.4in]{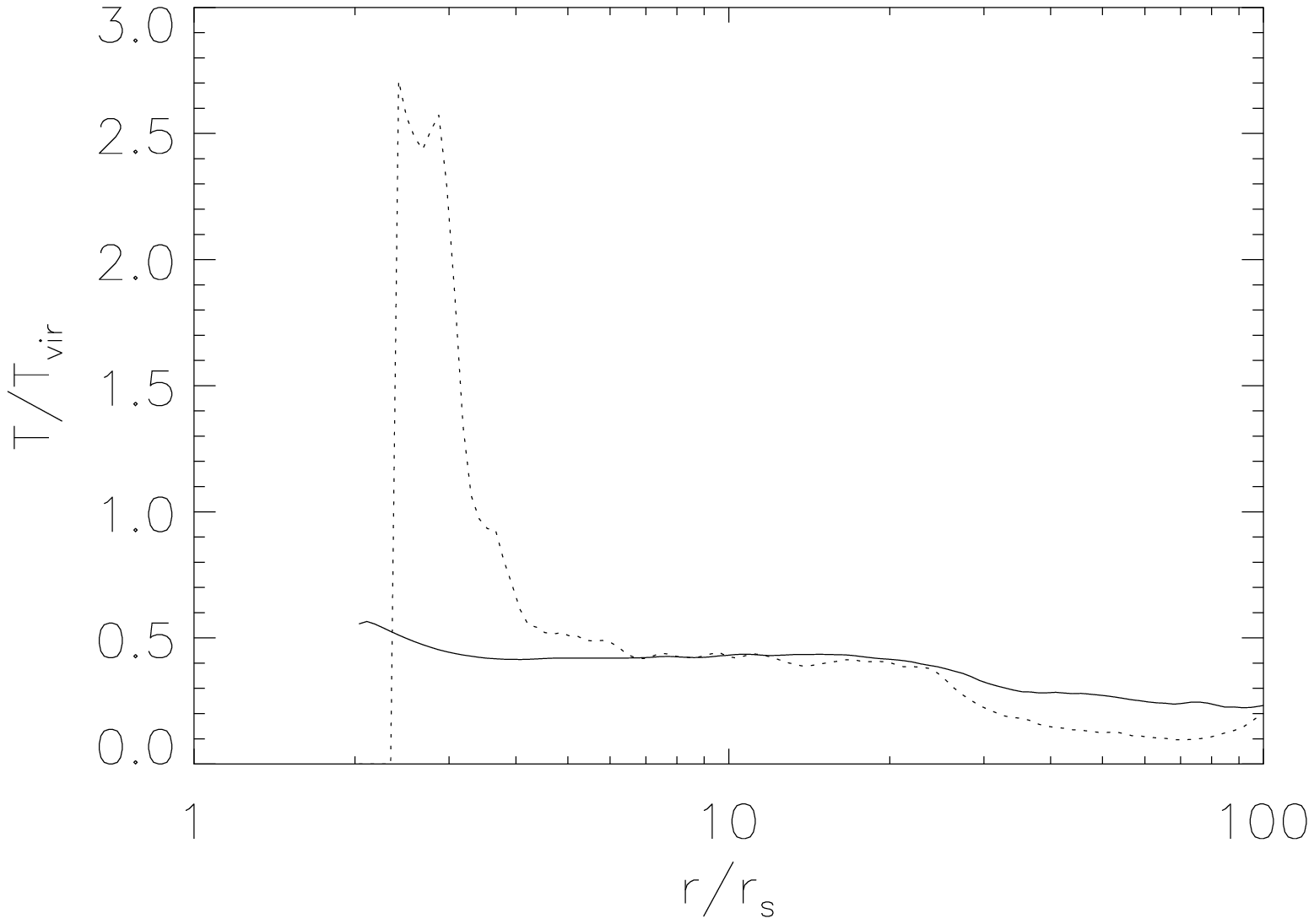}
 \caption{The radial distribution of flux-weighted temperature for Model A (left) and B (right). The solid and dotted lines are for inflow and outflow, respectively.}
\end{center}
\end{figure}

\begin{figure}
\begin{center}
 \includegraphics[width=2.4in]{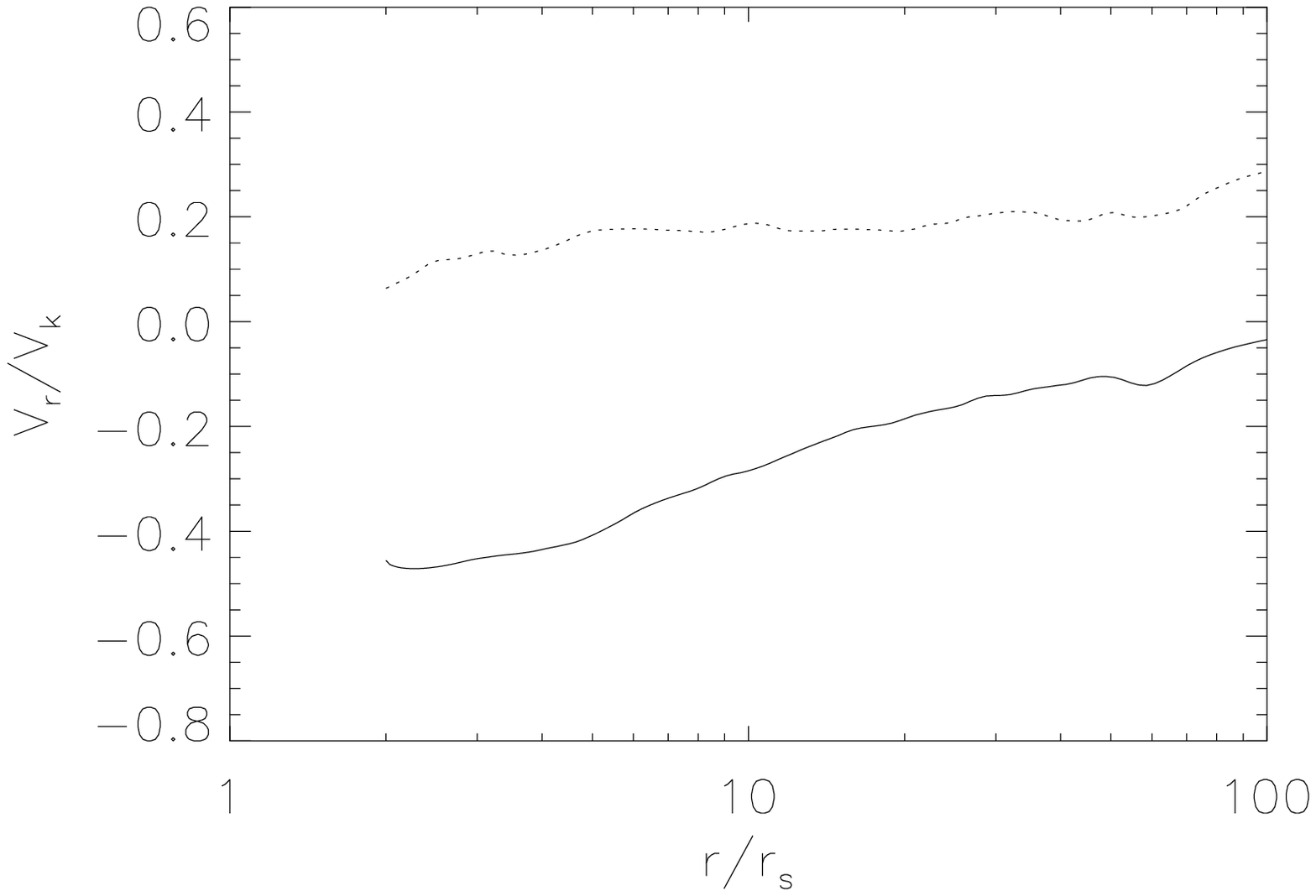}
 \includegraphics[width=2.4in]{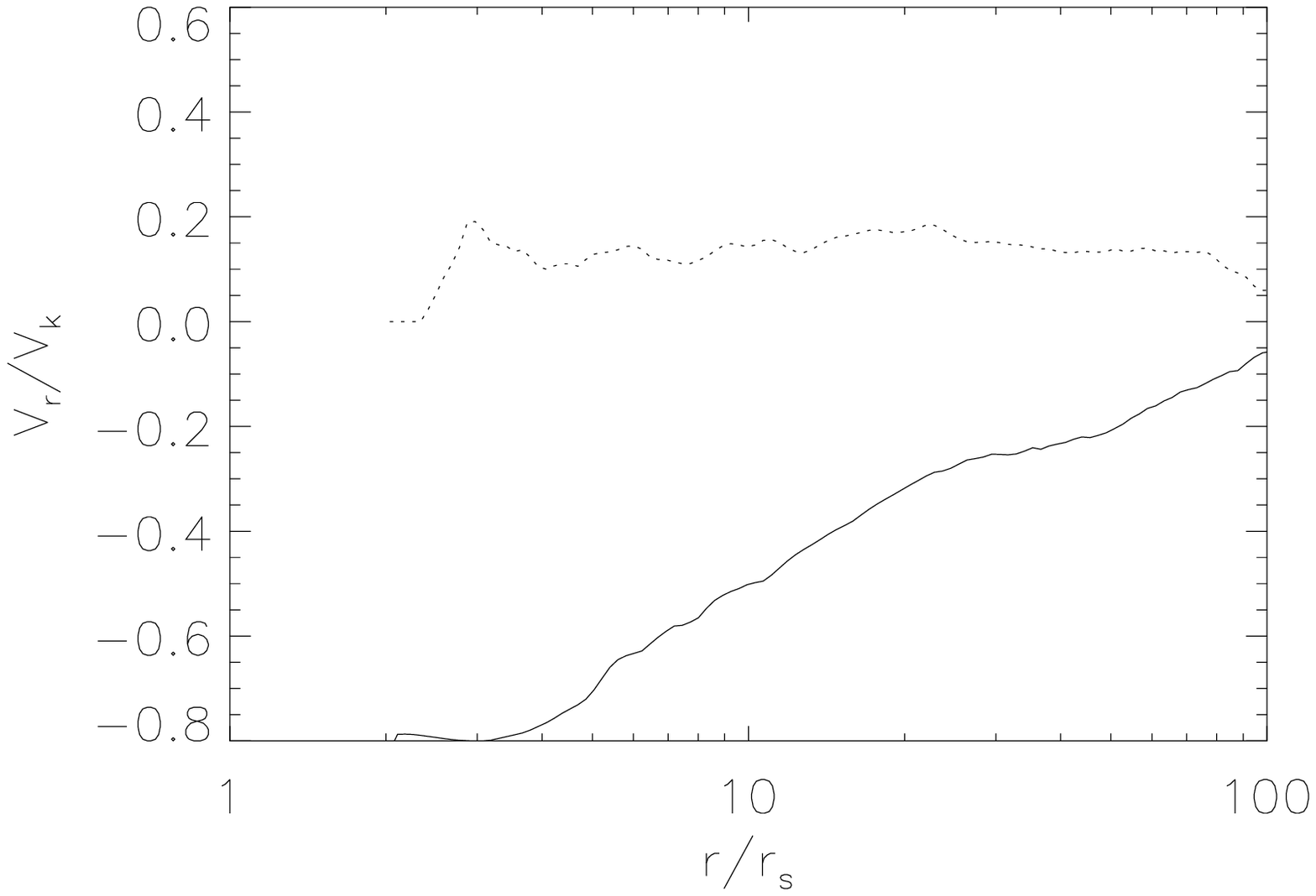}
\end{center}
 \caption{The radial distribution of the flux-weighted radial velocity for Model A (left) and B (right). The solid and dotted lines are for inflow and outflow, respectively.}
\end{figure}

\begin{figure}
\begin{center}
 \includegraphics[width=2.4in]{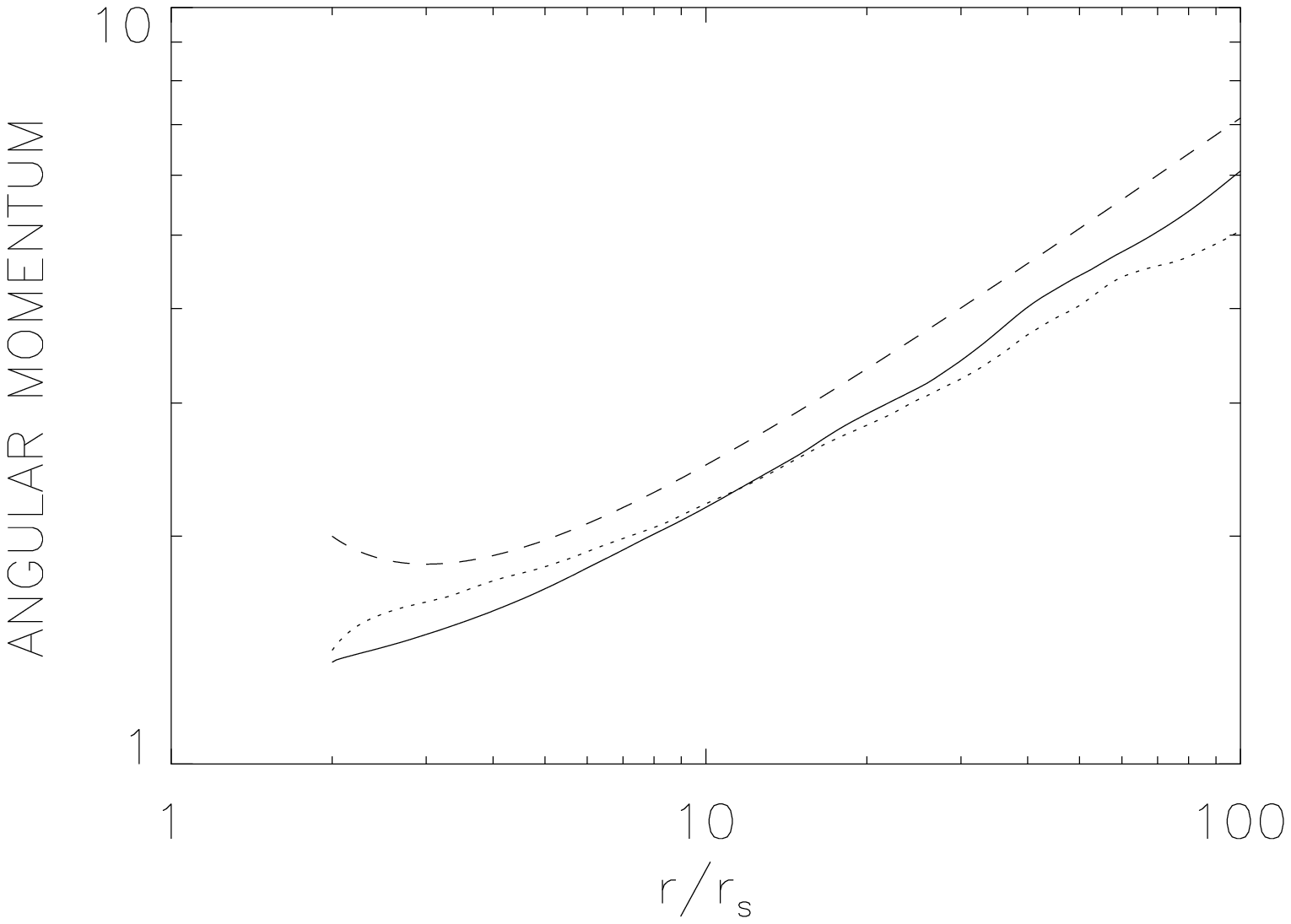}
 \includegraphics[width=2.4in]{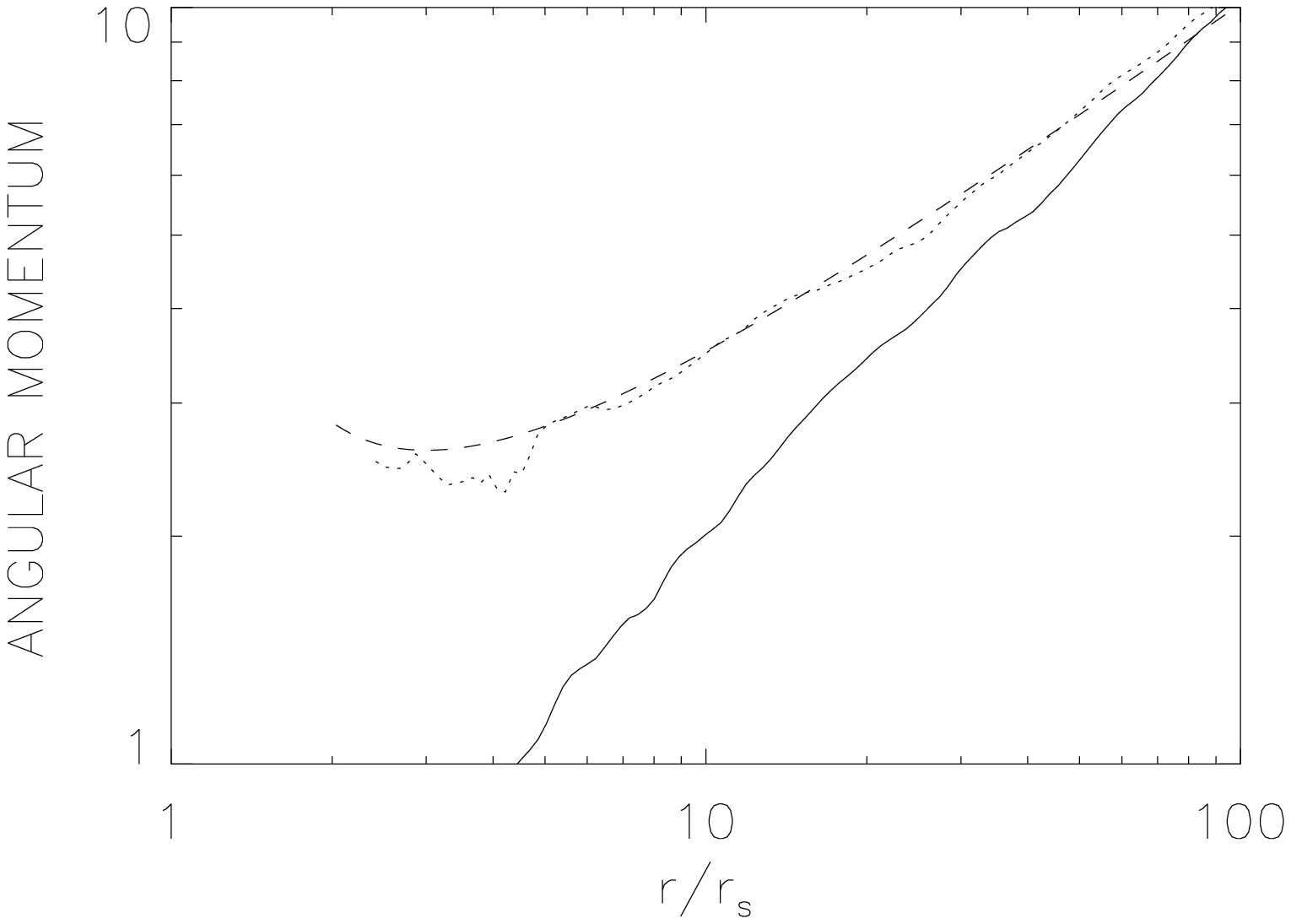}
 \caption{The radial distribution of the specific angular momentum $l=v_{\phi}r \sin\theta$ for Model A (left) and B (right). The solid and dotted lines are for inflow and outflow, respectively. The dashed line denotes the Keplerian angular momentum {at the equatorial plane}. }
\end{center}
\end{figure}

\subsection{The convective stability of MHD accretion flows}

We have analyzed the convective stability of an MHD accretion flow based on the H{\o}iland criteria. We found that the accretion flow is convectively stable. This result indicates that CDAF model at least can't be applied to MHD flows.

\subsection{The Outflow Scenario and the Origin of Outflow in HD and MHD Flows}

{\bf The overall scenario} suggested by the results mentioned above is that the inward decrease of the inflow and outflow rates is because of the mass lost in outflow. Outflows occur throughout the accretion flow from almost any radius and they must escape out of the outer boundary of the accretion flow. But note that the existence of systematic outflow does not exclude the existence of turbulence. A snapshot figure of velocity vectors which indicate the coexistence of both systematic inward and outward motion and turbulence is presented in Yuan, Bu \& Wu (2012). A similar conclusion is reached in Li, Ostriker \& Sunayev (2012), but Narayan et al. (2012) present a different point of view.


{\bf Origin of outflow in HD flows} was originally suggested to be due to the positive Bernoulli parameter of ADAFs. However, many numerical simulations, such as Stone, Pringle \& Begelman (1999), found the existence of outflow, while the Bernoulli parameter is negative. On the other hand, from Fig. 1, we see that the temperature of inflow is significantly lower than that of outflow. This result, combined with the fact that HD accretion flows are convectively unstable, suggest that outflow is produced by buoyancy.

{\bf Origin of outflow in MHD flows} must not be because of the buoyancy since accretion flow is convectively stable (\S2.2). We see from Fig. 3 that the specific angular momentum of outflow is very close to the Keplerian value, much larger than that of inflow. This suggests that the outflow is produced by the centrifugal force. Consider two fluid elements located at two different radii in a differential rotating accretion flow. Magnetic stress transports the angular momentum from the inner fluid element to the outer one. Once the angular momentum of the outer element reaches nearly Keplerian value, the centrifugal force will be able to make the fluid element turn around and throw it outward. This mechanism is different from the Blandford \& Payne mechanism since here we don't need a large-scale open magnetic field. In fact, we find that the magnetic field in both the accretion flow and coronal region is tangled. We therefore call it a ``micro-Blandford \& Payne'' mechanism.

\section{Possible Applications}

Many important problems remain to be probed. It is even not very clear in what condition the outflow can reach infinity, although they are able to escape out of the outer boundary of the accretion flow; and what is the exact values of outflow rate and terminal velocity. But still it is useful to discuss some possible observational applications which in turn can constrain theoretical models and supply some valuable clues. One is the Fermi bubble detected in the Galactic center. Our calculations indicate that the bubble can be readily inflated by the outflow from the ADAF (Mou et al. in preparation). Another example is the winds widely detected in AGNs and black hole X-ray binaries with various luminosities. It has been shown that in at least some sources a magnetic mechanism is required. People usually invoke Blandford \& Payne mechanism. However, it is still unclear how to realize the required large-scale open magnetic field in accretion flow. The micro-Blandford \& Payne mechanism is then a promising alternative. The readers are referred to Yuan, Bu \& Wu (2012) for details.

\end{document}